\documentclass[letterpaper]{article}
\usepackage{natbib,alifeconf}
\usepackage{url} 
\usepackage{amssymb}
\usepackage{hyperref}

\usepackage{listings}
\title{Rethinking Self-Replication: Detecting Distributed Selfhood in the Outlier Cellular Automaton}

\author{Arend Hintze$^{1,2}$ \and Clifford Bohm$^3$ \\
\mbox{}\\
$^1$Department for Data Analytics, Dalarna University, Sweden \\
$^2$BEACON Center for the Study of Evolution in Action, Michigan State University, United States of America \\
$^3$Michigan State University, United States of America \\
ahz@du.se, cliff.bohm@gmail.com}

\begin{document}
\maketitle

\begin{abstract}
Spontaneous self-replication in cellular automata has long been considered rare, with most known examples requiring careful design or artificial initialization. In this paper, we present formal, causal evidence that such replication can emerge unassisted --- and that it can do so in a distributed, multi-component form. Building on prior work identifying complex dynamics in the Outlier rule, we introduce a data-driven framework that reconstructs the full causal ancestry of patterns in a deterministic cellular automaton. This allows us to rigorously identify self-replicating structures via explicit causal lineages. Our results show definitively that self-replicators in the Outlier CA are not only spontaneous and robust, but are also often composed of multiple disjoint clusters working in coordination, raising questions about some conventional notions of individuality and replication in artificial life systems.
\end{abstract}


Data/Code available upon request

\section{Introduction}

In their work on the Outlier cellular automaton, \citet{yang2024emergence} describe the emergence of repeating structures suggestive of self-replication. In this paper, we investigate the basis of that replication using a formal, causal framework. Specifically, we show that the structures identified by \citet{yang2024emergence} meet a widely accepted criterion for genuine self-replicators: they produce multiple offspring that are causally traceable to a parent structure, and those offspring, in turn, generate further causally dependent descendants. Surprisingly, we find that the processes underlying replication often involve forking and merging causal pathways that coordinate to form distributed systems composed of multiple, spatially disjoint components. This fundamentally extends our understanding of how replication can emerge in complex systems. Our results and analysis go beyond Yang’s work, which reported recurring formations suggestive of replication but did not provide formal causal-lineage analysis or quantify generational depth and distributed multi-component structure.

Cellular automata (CAs) have a long and rich tradition in Artificial Life (ALife) research, serving as formal models to study complexity, emergence, and the fundamental nature of life-like processes. Some of the earliest inquiries in ALife revolved around the question, ``What is life?'' explored through CA systems \citep{langton1986studying}. While definitions of life vary, self-replication is widely accepted as essential~\citep{ganti2003principles,sipper1998fifty,robinson2005origins}.

Among the most influential early contributions, John von Neumann constructed a cellular automaton–based self-replicating machine, providing the first theoretical demonstration that self-replication could be achieved in a purely logical and computational system \citep{von1966theory}. His design required 29 distinct cell states to realize universal construction and self-replication, suggesting that significant complexity might be necessary for such phenomena. Von Neumann also introduced the concept of a universal constructor: a machine capable of constructing any arbitrary structure, including itself. Subsequent research sought simpler models of self-replication. Christopher Langton notably discovered a self-replicating loop requiring only eight states \citep{langton1984self}. Langton relaxed the criterion of universality, arguing that a self-replicator need only direct the reproduction of itself, not arbitrary structures. His ``Langton Loop'' exemplifies this idea: a continuous, cohesive configuration of cells operating as a single unit, using cycling internal instructions to orchestrate its self-replication. These foundational studies established the classical view of self-replication in CAs: a self-contained, connected entity that actively manages its own reproduction.

Langton's concept of self-replication was later extended by researchers who sought to explore not just simpler self-replicators, but also systems where such structures could evolve through interaction. Notably, the ``evo-loop'' system introduced by Sayama demonstrated that simple self-replicating loops could undergo evolutionary dynamics, including mutation and competition, within a deterministic CA framework \citep{sayama1999new,sayama2024self}.

\begin{figure}[b] 
    \centering
    \includegraphics[width=0.3\textwidth]{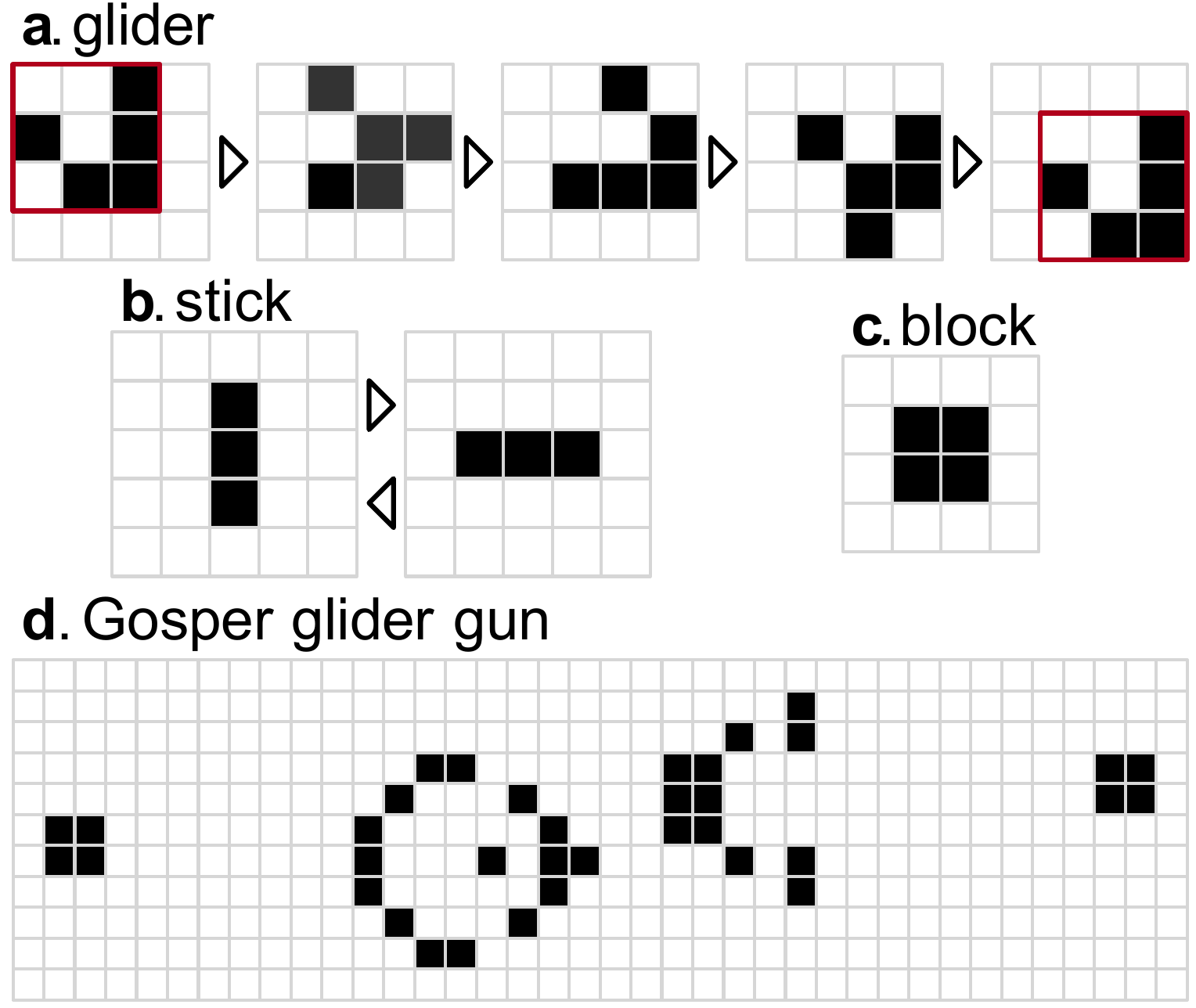}
    \caption{Various prominent patterns in the GoL CA. \textbf{a}. four steps of a glider leading to a repetitive and moving pattern. \textbf{b}. a 2-step oscillating stick pattern. \textbf{c}. a static block pattern. \textbf{d}. the Gosper glider gun.}
   \label{fig:glider}
\end{figure}

Alongside efforts to engineer explicit self-replicators such as Langton's loop, simpler and more minimalist CA were also being explored. A prominent example is Conway’s Game of Life (GoL) \citep{games1970fantastic}, a binary‐state CA in which each grid cell can be in one of two states, ``alive'' or ``dead.''
Each cell’s \textit{next} state is determined by its own previous state and the previous states of its eight immediate neighbors.
Despite its simplicity, GoL captivated researchers and the public alike with its rich emergent behaviors, including moving structures like gliders, self-sustaining ``glider guns'' which produce gliders, and a variety of stable or oscillating patterns (see Figure~\ref{fig:glider}). 
However, while the Game of Life demonstrates how simple local interactions can give rise to remarkable complexity, it has not been shown to spontaneously produce self-replicating entities ``out of the box.'' Thus, for studies focused specifically on the spontaneous emergence of self-replication, GoL is not an ideal testbed.

While many cellular automata exhibit rich and persistent dynamics, like GoL, the spontaneous emergence of self-replicating structures remains rare. This rarity has made identifying such structures in CA that can self-replicate challenging, particularly when moving beyond the well-understood, deliberately engineered examples like von Neumann's or Langton's designs. Only a handful of spontaneously emerging self-replicators have been discovered to date \citep{sapin2007research,sayama2009swarm}, and among the most notable recent examples is the ``Outlier rule,'' a binary-state CA exhibiting complex, self-replicating dynamics without explicit design \citep{yang2024emergence}.

Traditionally, as exemplified by Langton's loop and by natural biological organisms, self-replicators are conceived as cohesive, self-contained entities: discrete wholes that reproduce by making identifiable copies of themselves (allowing that mutations or developmental perturbations may occur during the process). In contrast, self-replication, in the Outlier CA, appears to involve aggregates of patterns — dynamic, shapeshifting groups of detached cells rather than static, unified structures.
This observation raises important questions about how individuality and replication should be understood in cellular automata. Before addressing these issues in detail, we must establish clear definitions for key concepts relating to patterns, alive versus dead states, and connectivity within these systems.

\subsection{Defining Terms}

While the alive or dead state of an individual cell is unambiguously defined in GoL, it is worth briefly justifying this interpretation. In GoL, a field of ``dead'' (off) cells will remain inert unless influenced by adjacent ``alive'' (on) cells. In other words, life begets life: dead cells do not spontaneously become alive without external influence. This dynamic aligns with a naturalistic interpretation and supports a further notion of alive cells as matter occupying space and dead cells as empty space. We note, however, that other CA rule sets exist in which spontaneous activation of dead cells can occur. In such cases, the intuition that alive cells represent matter and dead cells represent absence might fail. Nevertheless, this convention remains appropriate for the systems considered in this work and provides a consistent foundation for our analysis. 

We have been using the term ``pattern'' somewhat indiscriminately to refer to particular configurations of alive cells — structured collections that may or may not be contiguous or persist over time. We now introduce two formal types of patterns. Borrowing from~\cite{yang2024emergence}, we define a \textbf{cluster} as a collection of alive cells connected via Moore neighborhood adjacency (i.e., one can traverse between any two cells in the cluster without crossing a dead cell). We define a formation as any collection of clusters. For example, the collection of clusters that make up all elements on a CA grid is a formation, as is any subset of these clusters.

While the concepts of alive, dead, clusters, and formations describe the configuration of cells at a single time step in GoL, they do not allow us to describe its dynamic nature where identifiable patterns appear to emerge, persist, change, or even move over time. In the vernacular, ``an entity'' is defined as ``a thing with distinct and independent existence'' and we will adapt this definition to apply to CA. Formally, we define an entity in a CA, as an evolving set of cell states where each step depends causally on prior cell states of that entity, without external influence. Cluster entities form a subclass of entities, defined by the requirement that each temporal step consists of exactly one cluster. A glider (shown in Figure~\ref{fig:glider}.a), for example, is a cluster entity. However, entities need not be limited to single clusters: they may be formations that include multiple disjoint or interacting clusters, such as a glider gun, (shown in Figure~\ref{fig:glider}.d).

Entities can be further categorized by three pairs of traits. temporary versus permanent, static versus dynamic, and cyclic versus aperiodic. A permanent entity persists indefinitely when isolated (e.g., on an infinite grid), while a temporary entity terminates either inherently (e.g., collapsing into dead cells) or due to external disruptions (e.g., collisions). Static entities, like the $2 \times 2$ block in Figure~\ref{fig:glider}.c, have a fixed pattern and location over time. Dynamic entities, like gliders, change or move over time. A cyclic entity revisits an identical state (configuration of alive cells) after a finite period $t$, although the position may change. Cyclic entities include gliders ($t=4)$ the stick entity in Figure~\ref{fig:glider}.b ($t=2)$ and, the $2 \times 2$ block ($t=1$). Finally, an Aperiodic entity visits at least one state that is never revisited.

Even with these definitions, edge cases and ambiguities persist. For example, one could treat an entire instance of GoL --- the starting condition and all resulting world states --- as a single entity. This may seem counterintuitive, but because GoL is a closed system, all future states derive from the initial condition. In this sense, the whole system can be considered a single entity, and doing so does not preclude identifying smaller entities within it.
Another example: should an entity be considered cyclic or aperiodic if it eventually returns to a prior—but not original—state? In such a case, the entity is aperiodic, since some states are never revisited. However, if we define an entity only over the cycling period, then that entity is cyclic.
A more contentious question: is a glider gun a separate entity from the gliders it emits? We suggest that it is equally valid to view any evolution from a prior state as a single entity or as multiple entities — particularly when the resulting structures proceed with distinct and independent existences, as is generally the case with glider guns and gliders.
These questions all relate to \textit{individuation}, a concept widely explored in both philosophy and biology. In particular, \cite{hull1980individuality} and \cite{godfrey2009darwinian} argue that individuality is not binary, but graded and context-dependent. Further discussion is beyond the scope of this work, but we note that their frameworks emphasize causal cohesion, replication capacity, and evolutionary potential as key dimensions. This perspective aligns with the view advanced here: that multiple disjoint clusters may together form a replicating entity, and that individuation depends in part on how causally self-contained and reproductively autonomous a system is.

\begin{figure}[t] 
    \centering
    \includegraphics[width=0.5\textwidth]{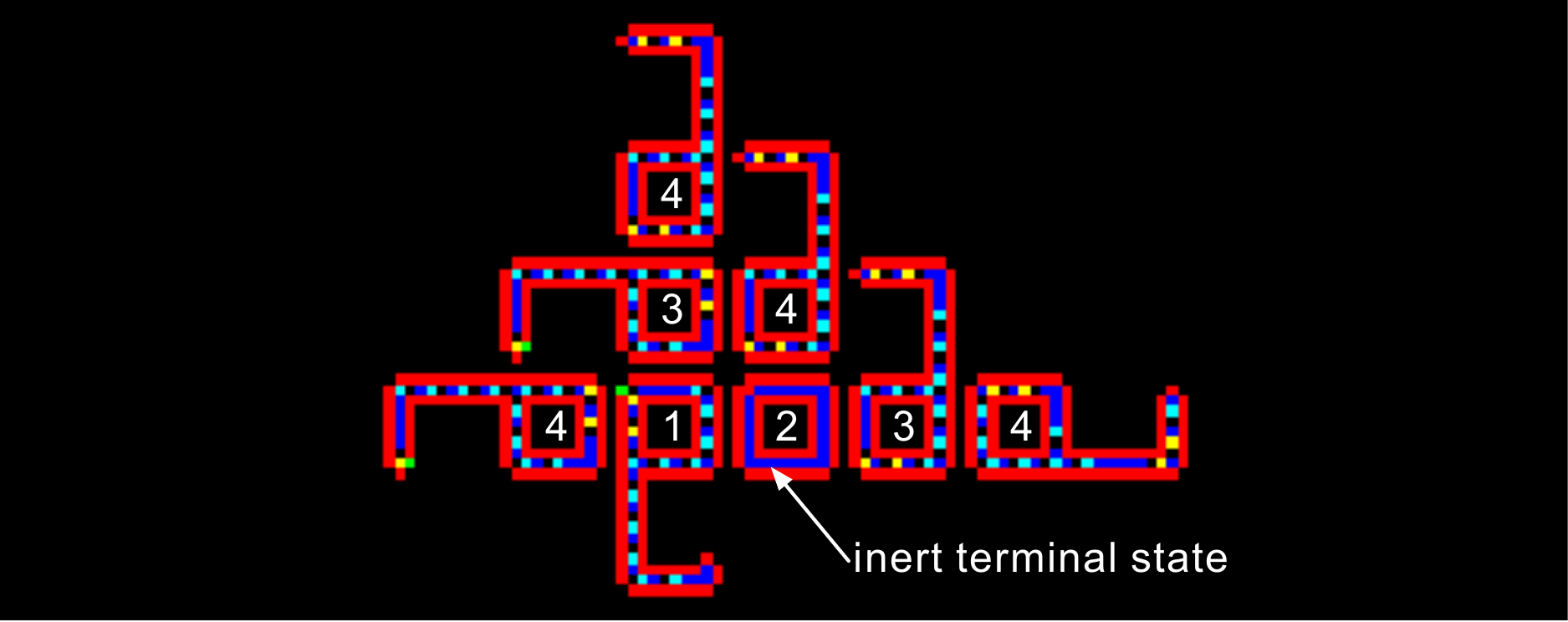}
    \caption{Illustration of Langton’s self-replicating loop. Loops, which may appear simultaneously, are labeled by order of appearance.
    }
   \label{fig:langton}
\end{figure}

\subsection {On Individuality and Replication}
Self-replication and its exact definition have been at the center of much debate.  A fundamental principle is that an individual (i.e., an entity) undergoes a transformative process that results in a copy that is also an independent individual (i.e., a new entity). With this in mind, we define a cluster as a \textit{self-replicating cluster} if its evolution can, in principle, produce at least two exact copies, each of which can be directly traced back to the original cluster. This definition requires that each offspring cluster is causally dependent on the original parent cluster, and not on any other offspring.
A glider does not satisfy this condition because each subsequent copy depends causally on the previous one, forming a single linear chain of descent rather than a branching phylogeny of independent offspring.
A glider gun is not a self-replicator as it produces copies of gliders, not of itself. To our knowledge, no self-replicating clusters have been identified in GoL. There are, however, informal reports of self-replicating formations, but these do not appear to have been published in formal literature.

Beyond classification, we informally define replication effectiveness as a cluster’s relative ability to reproduce reliably, frequently, and across varying conditions, providing terminology for later comparisons.

Our definition can be extended to apply to formations (i.e., a collection of disjoint clusters). In addition, our definition could also be amended to include approximate copies to allow for mutations. However, in this work, for technical reasons, we only consider clusters and perfect copies\footnote{This is a departure from \cite{yang2024emergence}, which defines a cluster to include rotational variants}.

Finally, we formalize what we mean when we use the general term \textit{replicator}. We have defined a self-replicating cluster and self-replicating formation as patterns that evolve in such a way that they result in derived copies. When we use the term \textit{self-replicating entity}, or simply \textit{self-replicator}, we refer to an entity --- the set of patterns that evolves over time --- that instantiates replication.

We return for a moment to Langton’s Loop to illustrate a key insight. Even under fixed and deterministic rules, replication outcomes depend on environmental context. Loop 2 (the second loop to develop in Figure~\ref{fig:langton}) is governed by the same update rules as its parent (Loop 1), yet it fails to replicate four times, as it's parent does, because it encounters a neighboring structure (in the form of it's parent) that constrains its replication.
Replication is not solely a property of an individual, but also depends on the state of the environment. In this case, loops require empty space to grow. The same replicator may succeed or fail depending on whether its surroundings support the conditions necessary for replication. Depending on the context, replication may require access to energy or resources, or other exacting conditions. Humans, for example, are not effective replicators when food is scarce, and we will fail to replicate entirely if we are not in an oxygen-rich environment!

In summary, an individual need not be a cluster (a connected collection of alive cells) nor must it be a static entity. In the next sections, we will discuss how, in the Outlier CA, replication often involves disjoint, spatially separated clusters that nonetheless function together as a cohesive unit, compelling us to reconsider what it means for something to count as an individual. The idea that multiple disjoint evolving clusters can comprise a replicating entity stems from~\cite{yang2024emergence} but, to our knowledge, has not been explicitly quantified using the methods described in the following section. 

\section{The Outlier Rule}

\begin{figure}
    \centering
    \includegraphics[width=0.45\textwidth]{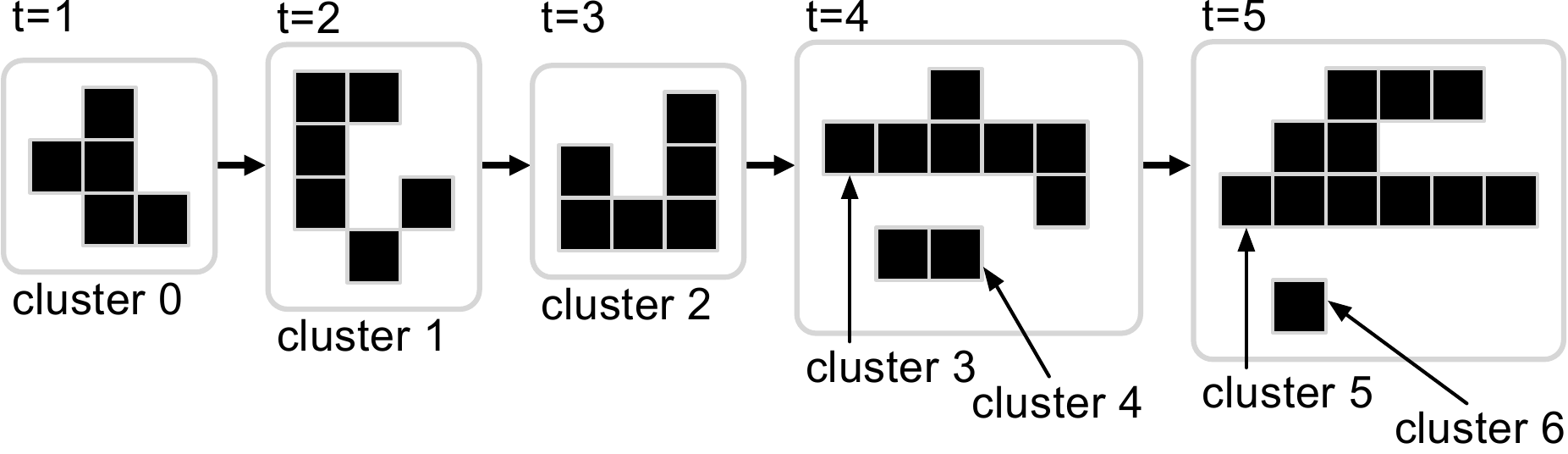}
    \caption{The sequence of the first five ticks of the Outlier rule CA when initialized with c0.}
    \label{fig:firstFive}
\end{figure}

To test our framework for identifying and analyzing self-replication in cellular automata, we required a CA with a rule set capable of producing complex, emergent behaviors, including the potential for spontaneous replication. While our methods are rule-agnostic, requiring only deterministic transitions and binary cell states, and so can be applied to any CA meeting these criteria, we chose the recently discovered Outlier rule as a testbed due to its uniquely rich dynamics and documented multi-scale pattern formation.

Our causal framework is rule-agnostic, requiring only deterministic transitions and binary cell states. It can thus be applied to any CA meeting these criteria.

The Outlier rule is a binary CA transition function documented in~\cite{yang2024emergence} that operates on a two-dimensional grid. Unlike more classic CA rules, such as Conway’s Game of Life, the Outlier rule was not hand-designed but evolved via novelty search, prioritizing behavioral diversity and the potential for open-ended evolution. Among hundreds of thousands of rules evaluated, the Outlier rule was notable for producing recurrent structures from sparse, unstructured initial conditions. The Outlier rule stands out among CAs for its ability to support multi-scale structures without engineered initial states or multistate logic, making it a compelling platform for studying distributed individuality and the dynamics of replication in minimalist systems.

The Outlier rule is a mathematical function that takes nine inputs --- the cell states of a $3 \times 3$ neighborhood --- and returns the state of the center cell on the next update. The function contains 271 steps, each executing either AND or XOR on a pair of inputs or the results of prior steps. Although internally represented as a multi-step logic, the function can be fully described as a static mapping between all 512 $3 \times 3$ binary neighborhood states and their resulting center cell states. To be clear, the algorithmic description used to generate the outlier rule is simply an artifact of its discovery process; the outlier rule is no more complex than any other explicitly defined binary CA acting on $3\times3$ neighborhoods. 

The full description of the outlier rule is available in Appendix 1 of~\cite{yang2024emergence}. The outlier rule in action can be seen at \href{https://lazyslug.com/lifeviewer/}{lazyslug.com/lifeviewer/} by setting the RLE data to:

{\ttfamily
\noindent{line 1 (remove line breaks):}

\noindent\detokenize{x=3,y=3,rule=MAPERETQB4eHWkQ7xD4eYZos}

\noindent\detokenize{BQZFixOBHmtFeehExrKVhURLRAqGxeIlSO1JY}

\noindent\detokenize{ZP6DRi69rop7TQCkvWTIag7kAS8g}

\noindent{line 2:}

\noindent\detokenize{bo$3o$2bo!}
}

In this study, we performed our analysis using a $1024 \times 1024$ grid with periodic boundary conditions. We initialized the system with the same seed cluster (c0) that was used in~\cite{yang2024emergence} as this is a simple pattern that does not result in early extinction. Figure~\ref{fig:firstFive} shows the first five time steps (ticks) of the outlier rules evolution, starting with c0.
The CA was run for 20,000 ticks. Around tick 10,000, alive cells crossed the periodic boundary, eliminating the previously uninterrupted leading edge. Before this transition, two distinct dynamic regimes were present: (1) the leading edge region, where the CA expanded into empty space, and (2) the internal region, behind the leading edge, where interaction among patterns resulted in more chaotic behavior. After approximately 10,000 ticks, only the interior region with its more chaotic dynamics remained.

\section{Method to Identify and Validate Replicators}
We define a self-replicator as an entity that produces at least two copies of itself, where each offspring is causally linked to the parent but not to each other. To quantify a self-replicator, we must provide full causal traces from a parent to each of its direct offspring. We chose to consider clusters as potential parents to limit the search space, but could have focused on formations instead. This requires a method for identifying clusters over time and reconstructing their causal lineage within the CA.

To achieve this, we construct a causal ancestry graph that captures how clusters give rise to other clusters over time. Recall that a cluster is defined as a set of adjacent alive cells connected via Moore neighborhood adjacency. We identify and track these clusters at each time step. When a new cluster is observed, it is assigned a unique cluster ID (CID). Cluster instances, specific occurrences at a particular time and location, are recorded using a unique identifier (UID) composed of the CID with a timestamp and the position of the leftmost alive cell in the cluster's top row.

Causation is traced at the level of individual cells. 
For each alive cell, we examine its $3 \times 3$ neighborhood in the previous time step. 
Because the Outlier rule is deterministic and fully specified for 
all 512 neighborhood configurations, we can determine which neighborhood cells caused each activation. We identify which specific neighboring cells were necessary for the transition. This allows us to trace each new alive cell back to the cluster or clusters that contributed causally to its activation. When aggregated to the cluster level, a cluster is causally derived from another if at least one of its cells depends on that ancestor cluster.

To link this information at the cluster level, we associate each contributing cell with the cluster it belonged to in the previous step, which we refer to as an ancestor cluster. If any cell in a new cluster depends causally on a cell from an earlier cluster, we draw a directed edge from the earlier cluster instance to the new one on the causal ancestry graph. In this way, we build a graph where nodes represent cluster instances at specific times and locations, and edges represent causal influence derived from the underlying update rules of the CA.

\begin{figure}
    \centering
    \includegraphics[width=0.125\textwidth]{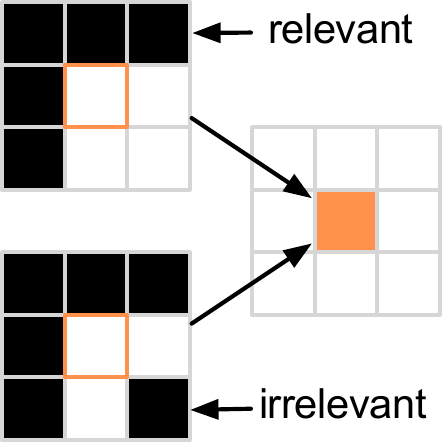}
    \caption{Illustration of a redundant update rule configuration. Whether the lone alive cell on the bottom right is alive or dead, the center cell will be alive on the next update. This redundancy highlights that not all adjacent ‘on’ cells are required, allowing the algorithm to exclude non-contributing cells when inferring ancestral relationships.}
    \label{fig:redundancy}
\end{figure}

Some update rules contain redundancy — not all cells in a $3 \times 3$ neighborhood are necessary to establish the state of the center cell. To avoid spurious causal links, we identify the minimal subset of neighbors that contribute to each new alive cell and exclude non-essential cells from the causal trace. This is illustrated in Figure~\ref{fig:redundancy}, where the bottom-right cell can be ignored as it does not affect the outcome. In rare cases where multiple minimal causal sets exist, our method conservatively includes all contributing clusters. While we found no such cases in the Outlier rule, our validation method is designed to handle them, supporting generalization to other rule sets.

With our cluster-level causal ancestry graph, it is trivial to detect self-replicating clusters; that is, to determine whether a given cluster results in multiple distinct copies of itself, each causally dependent on the original but not on one another.

\section{Analysis}
\begin{figure}
    \centering
    \includegraphics[width=0.27\textwidth]{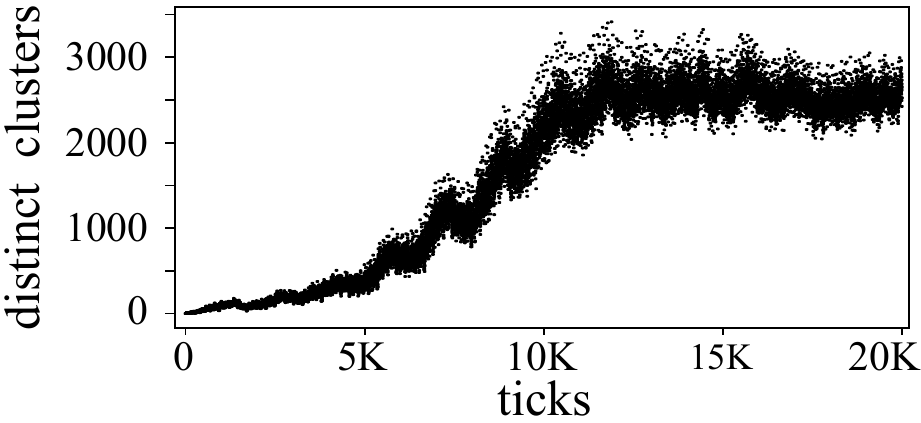}
    \caption{Number of distinct clusters observed over time in the Outlier CA. The x-axis represents time steps (ticks), while the y-axis shows the count of unique clusters present at each time step.}
    \label{fig:patternsOverTime}
\end{figure}

Applying our method to the Outlier rule over a 20,000-tick run on a $1024 \times 1024$ grid initialized with the c0 cluster produced a causal ancestry graph containing 31,959,320 nodes and 65,552,995 directed edges. Each node corresponds to a unique cluster instance — that is, a specific occurrence of a cluster at a particular time and position, identified by its UID. Edges indicate causal relationships between nodes. Across the full run, 966,208 unique clusters (novel contiguous arrangements of alive cells) were observed. As shown in Figure~\ref{fig:patternsOverTime}, the number of distinct clusters present at each time step grows approximately exponentially until around tick 10,000, at which point the system’s leading edge reaches the grid’s periodic boundary.

\subsection{Replicator Classification}
In their analysis of the Outlier rule, \citet{yang2024emergence} suggests that recurring formations constitute self-replicating entities. Here, we provide formal evidence to support that claim. Leveraging the causal ancestry graph, we can definitively determine whether self-replicators are present in the Outlier rule. Our results confirm that they are. For tractability, we focus our analysis on clusters rather than formations. As we demonstrate, clusters provide sufficient causal evidence of self-replication, and their replication implies the replication of the larger formations in which they participate.

We first consider the original seed cluster, c0. Within the first 10,000 ticks, we observe 433 copies of c0, each of which can be causally traced back to the original c0 instance. However, none of these offspring go on to produce additional c0 clusters within that period. Thus, while c0 qualifies as a self-replicating cluster under our definition, it is a relatively ineffective one, exhibiting only a single generation of replication.

In contrast, the second cluster to appear, c1, is a more robust self-replicator. Over the same 10,000-tick interval, c1 produces 1,677 offspring, all causally linked to the original c1 instance. Near tick 10,000, we observe one c1 offspring producing three additional c1 clusters, demonstrating a second generation of replication. We hypothesize that the first-generation offspring of both c0 and c1 are primarily the result of the expanding leading edge, while the emergence of second-generation offspring may be due to interactions at the boundary where opposing wavefronts meet following the system’s periodic closure.

\begin{figure}[t]
    \centering
    \includegraphics[width=0.45\textwidth]{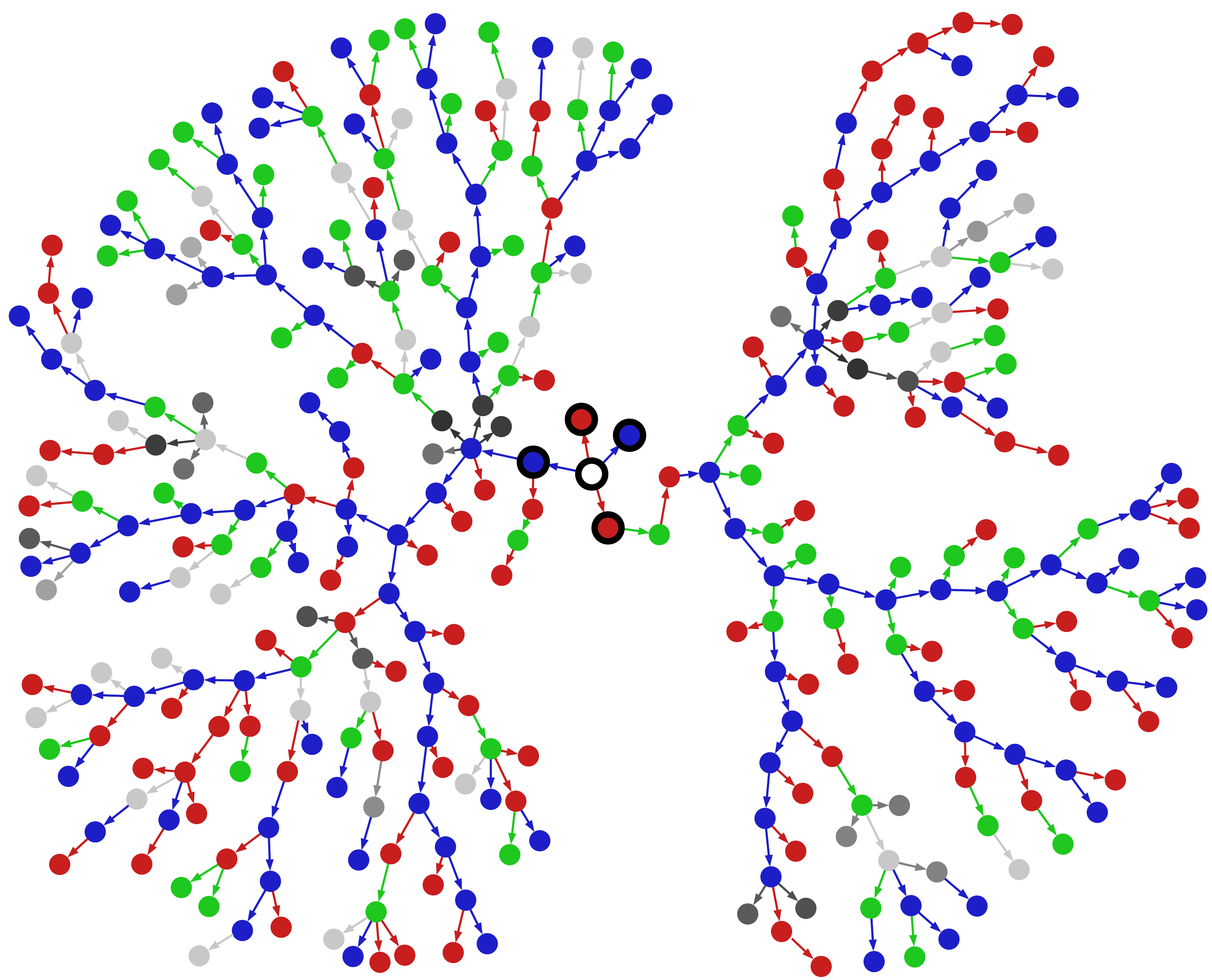}
    \caption{Phylogenetic tree of self-replicators derived from the initial c2. Node and link colors indicate the replication times. the initial c2, in white, and each of the initial four offspring shown in Figure~\ref{fig:fourBranches} are outlined in black}
    \label{fig:phylogeny}
\end{figure}

The third observed cluster, c2, is the first to exhibit robust, sustained self-replication. Over the first 10,000 ticks, c2 appears 2,439 times—replicating more frequently than either c0 or c1. Crucially, a substantial number of c2 offspring go on to reproduce themselves. Figure~\ref{fig:phylogeny} shows the phylogenetic tree rooted at the original c2 instance. For clarity, the figure includes only the 344 instances that produced at least one offspring. Over the 10,000-tick interval, we observe 15 generations of c2 replication, providing clear evidence of a lineage of persistent, self-replicating clusters.

\begin{figure*}[t]
    \centering
    \includegraphics[width=1\textwidth]{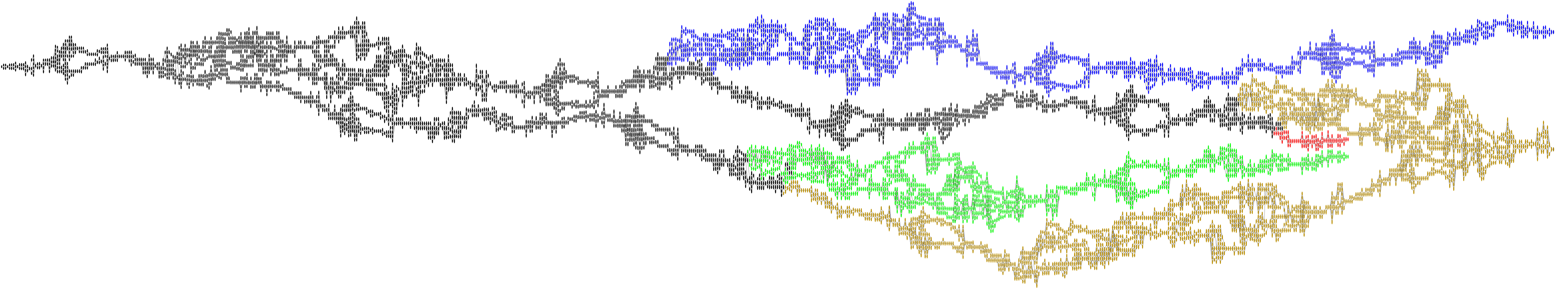}
    \caption{Subgraph of the complete causal pathways resulting in four offspring from the initial c2 cluster. Time and causal progression proceed from left to right. Each node represents a specific cluster instance, and edges indicate causal dependency (each node is fully determined by the connected nodes to its left). Red, green, orange, and blue nodes indicate clusters that occur exclusively within the causal pathway of one offspring, with the rightmost terminal nodes being the replicated c2 instances. Black nodes represent clusters that appear in the causal pathway of more than one offspring.}
    \label{fig:fourBranches}
\end{figure*}

The causal ancestry graph produced by our method represents a different kind of lineage than a traditional organismal phylogeny. Rather than showing evolutionary relationships between organisms, it captures the step-by-step causal descent of cluster instances — individual occurrences of specific patterns across space and time. We refer to the traced causal lineage of cluster instances leading to a replicator as a \textit{developmental cluster phylogeny}. 

Due to the scale of the full causal ancestry graph, we cannot visualize the entire structure, showing relationships among all clusters. Instead, in Figure~\ref{fig:fourBranches}, we present a focused subset: the developmental cluster phylogeny of the first four c2 offspring that each go on to replicate. This plot traces every intermediate cluster instance in the lineage connecting the original c2 to each of these four replicating descendants. The four branches are color-coded (red, green, blue, and orange) to denote developmental paths associated with only one of the four offspring. Segments shown in black represent nodes that are part of more than one causal pathway; clusters that contribute to the development of multiple offspring. Notably, during the final 143 ticks of their respective developmental trajectories, the red, green, and blue branches pass through an identical sequence of clusters, suggesting a shared late-stage developmental trajectory. In contrast, the orange branch follows a distinct developmental pathway up until the final step, where the c2 pattern reappears.

\begin{figure}[h]
    \centering
    \includegraphics[width=.47\textwidth]{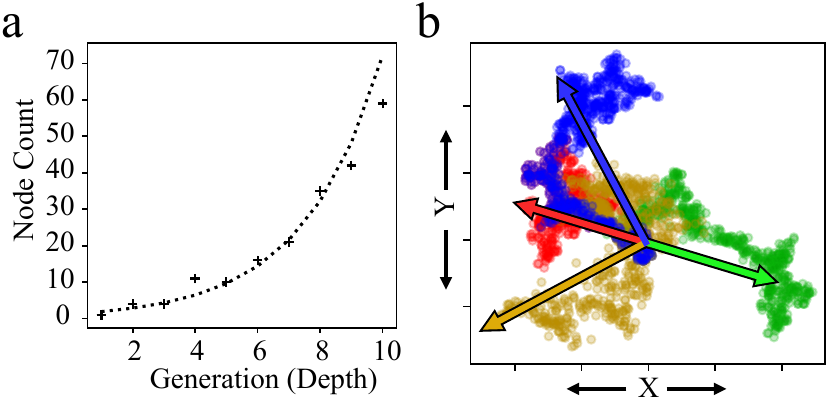}
    \caption{(\textbf{a}) Exponential growth of self-replicators derived from the initial c2. The x-axis shows generation depth; the y-axis, number of replicators. Dashed line shows exponential fit with growth factor $1.4955$. 
    (\textbf{b}) Spatial extent of the initial c2’s replication process. Each point is the mean location of all clusters, associated with a given offspring, at a given tick within one of four branches (see Figure~\ref{fig:fourBranches}). Arrows mark start and end points; fast branches shown in red and green, slower ones in blue and yellow.}
    \label{fig:exponential_growth}
\end{figure}

To quantify the growth rate of c2, we measured the size of each generation and performed a linear regression on the logarithm of the node counts for the first 10 generations, plotted as a function of generation number (see Fig.~\ref{fig:exponential_growth}.a). Assuming an exponential growth model \( N(d) = r^d \), where \( r \) is the growth rate and \( d \) is the generation number, the best-fit line yielded a growth factor of approximately 1.4955. While this model suggests sustained exponential growth, spatial constraints begin to limit replication around tick 10,000, when the expanding front reaches the grid's periodic boundary, curtailing further expansion in our simulation.

Curiously, the original c2 instance does not divide into two immediate offspring. Instead, c2 produces many offspring, although only four of these successfully self-replicate. Among these four, two require 675 ticks to develop, while the other two take 778 ticks. This divergence raises an important question about the relationship between replication time and reproductive success. Intuitively, one might expect faster replication to result in greater long-term reproductive output. However, when examining the full phylogeny, we observe the opposite: the 675-tick lineage produces only 96 replication events, while the longer 778-tick lineage results in 125.

The explanation lies in how replication propagates through space. The two faster branches expand in approximately horizontal directions (see Figure~\ref{fig:exponential_growth}.\textbf{b}, red and green), while the longer-tick branches are oriented obliquely, offset by roughly 90 degrees from one another while also out of line with their faster siblings. As a result, descendants along the faster replication paths are more likely to collide, reducing lineage persistence. In contrast, the longer-tick branches experience less spatial interference and therefore sustain more successful replication events. This suggests that available space—and the spatial direction of replication—is a critical environmental factor that shapes reproductive success in this system.

We have discussed four c2 replicators, each following a distinct developmental trajectory. However, when analyzing all replication events in Figure~\ref{fig:phylogeny}, we were surprised to find far more diversity than expected. Rather than uncovering only variations of the four previously observed self-replication processes --- each completing in either 675 or 778 ticks --- we identified a total of 18 distinct replication times. While the 675 and 778-tick cycles are the most common, we also observed a 881-tick replication occurring 67 times, and a 572-tick cycle appearing 30 times. The remaining 14 replication durations occurred rarely, each fewer than four times.

Moreover, as seen in Figure~\ref{fig:patternsOverTime}, each of these replication processes corresponds to a distinct sequence of cluster states. Thus, the number of unique developmental pathways must exceed the number of observed replication times. A complete accounting of these pathways remains an open task for future work, but these results already demonstrate a surprising degree of variation in how even a single cluster type can replicate.

\subsection{Pattern Emergence in Unbounded Space}
To investigate spatial dynamics without boundary constraints, we ran the Outlier CA for 75,000 ticks on an effectively infinite grid, tracking only the positions of alive ($1$) cells. For each new cluster, we recorded its location, size, and time of appearance.

\begin{figure}[t]
    \centering
    \includegraphics[width=0.41\textwidth]{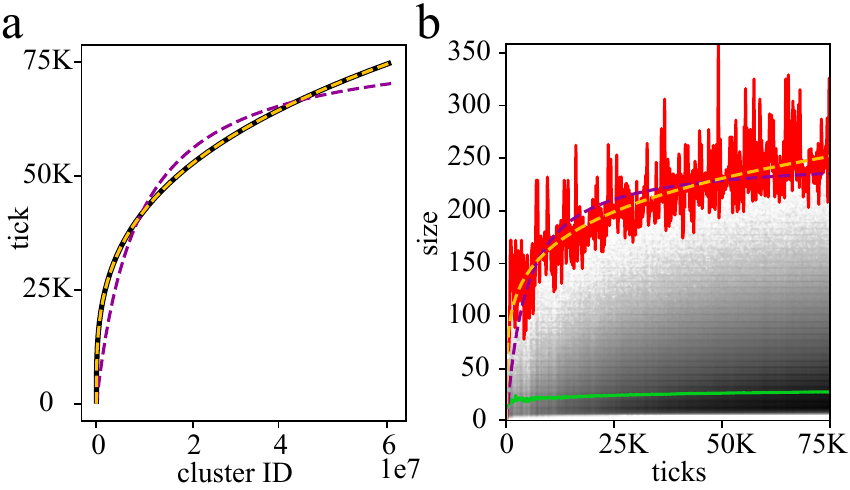}
    \caption{
    (\textbf{a}) Time of appearance (y-axis) for each novel cluster (x-axis), with power-law (dashed orange) and hyperbolic (dashed purple) fits. 
    (\textbf{b}) Maximal sizes of newly discovered clusters (500-tick bins on x-axis), with power-law (orange) and hyperbolic (purple) fits shown as dashed lines. The mean cluster size is plotted in green; the gray background shows the $\log_{10}$ density of newly found clusters. Note the horizontal bands, indicating overrepresentation of cluster sizes that are multiples of six..
    }
    \label{fig:heatmap}
\end{figure}

Three observations emerged (Figure~\ref{fig:heatmap}): (a) new clusters appeared predominantly within a diagonally expanding square region centered on the origin (not shown); (b) their emergence times followed a power-law distribution (AIC = 104114.5), better than a hyperbolic fit (AIC = 163790.6), suggesting ongoing cluster generation \cite{wiser2013long}; and (c) maximum cluster size also followed a power-law (AIC = 1949.6 vs. 2046.1), indicating indefinite growth in scale.

As new clusters emerge, particularly near the center, they frequently interfere with existing replicators, introducing structural variation. Like mutation in EvoLoop systems \cite{sayama2024self}, these interactions can produce novel replicators composed not of single clusters, but coordinated sets of clusters.

To explore this, we traced all descendants of the original c2 replicator from $t=3$, identifying 187 self-replicators. We then searched for additional c2-like replicators appearing after $t=5000$ and found 205, including 18 that do not descend from the original c2. These new replicators exhibit structural and temporal trajectories that differ substantially from their ancestor, suggesting a genuine expansion in replicator diversity within the system.

\section{Discussion}

\subsection{Distributed Replication}

Unlike engineered replicators such as Langton's loop, which generate offspring through a single, cohesive pattern that orchestrates replication from a central structure, the replicators we observe in the Outlier CA exhibit fundamentally different behavior. Its replication process unfolds through multiple spatially disjoint patterns, which remain disconnected for several ticks before either merging or branching further. This distributed behavior proves that replication can be composed of loosely coupled, cooperating components rather than a monolithic control structure.

While the initial replication processes (e.g., with durations of 675 or 778 ticks) represent some of the earliest observed pathways, there is no reason to view them as having privileged status. These lineages likely developed along the expanding leading edge, where interactions with other structures were minimal. As the system evolves and chaotic behavior behind the leading edge becomes more prevalent, additional interactions introduce further variation in developmental trajectories, not as disruptions of a baseline process, but as natural continuations of the same fundamental dynamics operating in a more complex environment.

Our analysis focuses on live cells and their causal dependencies. While every cell --- alive or dead --- has a causal history, we do not consider the histories that lead to cells being dead when tracing replication pathways. This is a deliberate modeling choice: in the Outlier rule, only alive cells can give rise to other alive cells. That is, life begets life, and the propagation of structure is carried exclusively by living cells. It may be tempting to argue that the absence of certain structures --- such as a cell that might have been alive but was not --- should factor into causal explanations of replication. However, this conflates the causal transmission of structure with the constraints that shape which pathways are realized. Our focus is on identifying what actively contributes to replication, not what might have occurred in counterfactual histories.

\section{Conclusion}
In this work, we have shown how spontaneous self-replicators emerge within a simple, binary CA; not as cohesive, hand-engineered artifacts, but as distributed systems composed of multiple, spatially disjoint components. Using a formal, causal tracing method, we demonstrated that these structures meet the criteria for self-replication: they produce multiple causally distinct offspring, which in turn produce further offspring, forming branching lineages of causal descent.

To our knowledge, Figure \ref{fig:fourBranches} presents the first complete description of a non-engineered, multi-component self-replicator in a 2D discrete CA or continuous CA such as Lenia \cite{chan2018lenia,chan2020lenia}. This finding extends the previous understanding of self-replicators, which were traditionally considered to be single-unit, cohesive, and self-contained structures ~\cite{von1966theory,langton1984self,sayama2024self}, and provides formal support for earlier informal observations by \citet{yang2024emergence}.

Extending these methods to other systems is critical to validating the generality of our findings. A natural next step would be to analyze Evoloops~\citep{sayama1999new}, a deterministic, multi-state CA where, initially engineered, self-replicators evolve through environmental interactions. Investigating the causation of replication and how distributed information relates to the emergence of new modes of replication could further clarify the role of causal cohesion in CA.


In the Outlier CA, replication occurs without any explicit separation of roles: the information and the mechanism are distributed across multiple interacting components. The sets of patterns that participate in self-replication contain the information needed for replication. This differs from many intuitive notions of replicators, such as von Neumann's universal constructor, which separated information (the tape) from the constructor mechanism. In the replicators we have analyzed, the replication logic is embedded in structure; however, it is neither centralized nor explicitly encoded.

The Outlier replicators we observe do not persist because they were designed to do so, nor because of stochastic fluctuations. They persist because their structure leads, under the governing rules, to the reappearance of similar structures. In this view, replication is not a directed process but a consequence of how certain configurations propagate themselves within the system’s dynamics.

This reframes the notion of replication. In dynamical systems, persistence need not be defined by the continued existence of a static structure, but can also arise from the recurrence of a stable causal process. In such cases, what persists is not any particular configuration, but the process by which such configurations are generated. In this view, the replicating entity is the process itself, the complete sequence of causal steps that sustains and re-creates itself. When replication occurs, what is reproduced is the process, not merely its transient outputs.

In a closed, deterministic system where all future states are the result of prior configurations, replication emerges as a natural and inevitable phenomenon, and once established, ensures persistence. It is not an anomaly requiring special construction, but a general consequence of how structure arises and is maintained in systems governed by local interactions and finite rules. From this perspective, replication is not merely a feature of life but a fundamental mode by which information, organization, and complexity endure.

\section*{Acknowledgments}
Large Language Models (ChatGPT, Claude, and DeepSeek) were used to refine, but not create, content in this work.

\footnotesize
\bibliographystyle{apalike}
\bibliography{sample}

\end{document}